\documentstyle[twoside,fleqn,espcrc2,psfig]{article}

   \newcommand{\Z}{{\cal Z}}
   \newcommand{\be}{\begin{equation}}
   \newcommand{\ee}{\end{equation}}
   \newcommand{\la}{\label}

\newcommand{\AmS}{{\protect\the\textfont2
  A\kern-.1667em\lower.5ex\hbox{M}\kern-.125emS}}

\title{Finite density QCD with heavy quarks}

\author{R. Aloisio\address{Dipartimento di Fisica, Universit\'a dell'Aquila, 
        via Vetoio, 67100 L'Aquila, Italy}$^{,d}$,
	V. Azcoiti\address{Departamento de F\'isica Te\'orica,
	Facultad de Ciencias,Universidad de Zaragoza,50009 Zaragoza,Spain}
	G. Di Carlo\address{Istituto Nazionale di Fisica Nucleare,
	Laboratori Nazionali di Frascati, P.O.B. 13,
	00044 Frascati, Italy},
	A. Galante$^b$\thanks{Talk presented by A. Galante},
	A.F. Grillo\address{Istituto Nazionale di Fisica Nucleare,
	Laboratori Nazionali del Gran Sasso, 
	67010 Assergi, Italy}}
 
\begin{document}
\pagestyle{empty}
\begin{abstract}
In the large fermion mass limit of QCD at finite density the
structure of the partition function greatly simplifies and can be studied
analytically. We show that, contrary to general wisdom, the phase of the Dirac 
determinant is relevant only at finite temperature and can be neglected 
for zero temperature fields. 
\end{abstract}

\maketitle
\section*{Introduction}
Finite density QCD is still a poorly understood field from an analytical point of view. 
We miss any quantitative data about the passage from the hadronic to the 
quark-gluon plasma phase at high densities and even more qualitative 
information like the order of the phase transition (if any) is not available.  

This situation mainly reflects the difficulties to perform numerical 
simulations of QCD 
at finite chemical potential $\mu$. 
Almost all the standard algorithms for zero density QCD include the 
fermionic determinant in the 
generation of configurations and fail for complex valued determinants.
Attempts to create ad hoc algorithms have been made but they are generally
very computer consuming, they rely on some a priori unjustified 
approximations
to get real valued observables and the results obtained up to now are plagued 
from unphysical effects like the early onset of the baryon density
\cite{bar1}, \cite{noi1}. 

The main difficulty is to evaluate correctly the contribution of the phase 
of the action. In fact the partition function  can be written as the product 
of two terms: 
\be\la{1}
\Z = \Z_\| \langle\cos\phi\rangle_\|
\ee
where the first is evaluated using the
modulus of the Dirac determinant and the second  is the mean value of 
the cosine of the phase calculated weighting the configurations
with the usual pure gauge part of the action times the modulus of the determinant.
In the thermodynamic limit $\langle\cos\phi\rangle_\|$
gives a finite contribution to the free energy density
if it is proportional to $e^{-V}$. 
If this is the case there is no chance to determine numerically its value:
for any reasonable statistic, is not possible to evaluate it from a set of 
measurements of $O(1)$.  

The simple idea to consider only the modulus of the action has
received very little attention, especially  after the studies on the random
matrix model where the phase of the Dirac determinant is
fundamental for obtaining the correct result \cite{step}.
In QCD the one dimensional case can be addressed analytically and we
can see that the average value of the cosine of the phase goes to 1 as
we approach the thermodynamic limit \cite{noi2};
the physically relevant four dimensional theory can only be studied in the large bare 
mass limit where several simplifications in the fermionic determinant 
structure occur.
\section*{The strong coupling limit}
We can first write the fermionic matrix $\Delta$ separating the contribution of
forward (backward) temporal links $G$ $(G^\dagger)$ and spatial links $V$:
\be\la{2}
2\Delta = 2 m I + e^\mu G + e^{-\mu} G^\dagger + V 
\ee
When $m\gg 1$ only the mass term and the forward temporal links can
contribute to the fermionic determinant. Since quarks do not propagate
$\det\Delta$ can be written as a function of traces of Polyakov loops 
($L_i$) only and we get, except for an irrelevant multiplicative constant,
\be\la{3}
\det\Delta = e^{3V_sL_t\mu} \prod_{i=1}^{V_s}( c^3 + c^2 Tr L_i +
c Tr L_i^* +1)
\ee
where $V_s$ is the spatial volume, $L_t$ the lattice temporal extent and 
$c=(\frac{2m}{e^\mu})^{L_t}$ \cite{tous}.

In the strong coupling limit we can solve exactly the theory since the
Polyakov lines are independent and the partition function factorizes
trivially:
\be\la{4}
\Z (\beta = 0) = e^{3V_sL_t\mu} \left( c^3 + 1 \right)^{V_s}
\ee
It follows immediately that, in the thermodynamic limit,
the system undergoes a first order transition at $\mu_c=\log 2m$ where
the baryon density jumps from zero to the saturation value.
Whatever the spatial volume is the singular behaviour in the free 
energy density is related to $L_t$ only.

Regarding equation (\ref{1}) we see that an upper bound for $\Z_\|$ 
is given by the partition function at $\beta=\infty$ and therefore
\be\la{5}
\langle\cos\phi\rangle_\| \geq \left(\frac{c^3 + 1}{(c + 1)^3}\right)^{V_s}
\ee
As the volume goes to infinity this ratio goes to 1 except when
$\mu = \mu_c$ where it goes to zero exponentially with the spatial volume; 
we have a correct description of our zero temperature system replacing the fermion 
determinant by its absolute value.
\begin{figure}[!t]
\psrotatefirst
\psfig{figure=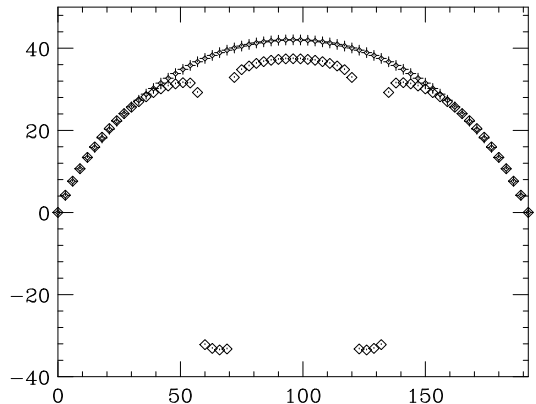,angle=90,width=200pt}
\caption{Logarithm of the modulus of the GCPF coefficients times
the sign of the coefficients as a function of the coefficient order: 
analytic (crosses) and numerical (diamonds) results for $V=4^3\times 4$.}
\label{fig:1}
\end{figure}

The situation changes completely if we consider the effect of non zero
temperature: we must keep $L_t$ finite and consider the limit of infinite 
spatial volume. From (\ref{4}) we see that the transition becomes a
smooth crossover and, from the definition (\ref{1}),
 we can also realize easily that $ \langle\cos\phi\rangle_\| \sim e^{-V_s}$.
Since the system volume is proportional to $V_s$ this implies that the
phase can not be neglected.

The availability of the analytical solution (\ref{4}) can also provide a
useful tool to check the results of numerical simulations.
Similarly to the Grand Canonical Partition Function formalism we have 
written the partition function $\Z$ as a polynomial in the 
variable $c$ and studied the convergence properties of  the coefficients.
We generated several thousands of random gauge configurations and calculated
the averaged expansion coefficients. This has been done for lattices $4^4$ 
(fig. 1), $4^3\times 20$ (fig. 2) and $10^3\times 4$ (fig. 3) and results 
are plotted, as a function of the coefficient order, superimposed to the 
ones obtained from (\ref{4}). 
From the plots we can realize that $i)$
numerical results are not positive definite as they should be but
for large $L_t$ this phenomena disappears;
$ii)$  even considering the modulus of the averaged coefficients
we do not get the correct order of magnitude in the central region;
$iii)$ increasing $L_t$ or $V_s$ the situation makes worse.

This seems to indicate that, in this model, our configuration
ensemble is not relevant for the physics at intermediate $c$ ($i.e.$ $c\sim 1$)
and a correct determination of all the coefficients from a random
ensemble is not possible. 
Increasing the temporal lattice extent the contribution of the phase
becomes smaller and, with reasonable statistic, we get positive averaged 
coefficients. Nevertheless their convergence to the correct values do not improve
suggesting that a non accurate determination of $\langle\cos\phi\rangle_\|$
may not be the only source of artifacts in simulations (see also \cite{bar1}).
\begin{figure}[!t]
\psrotatefirst
\psfig{figure=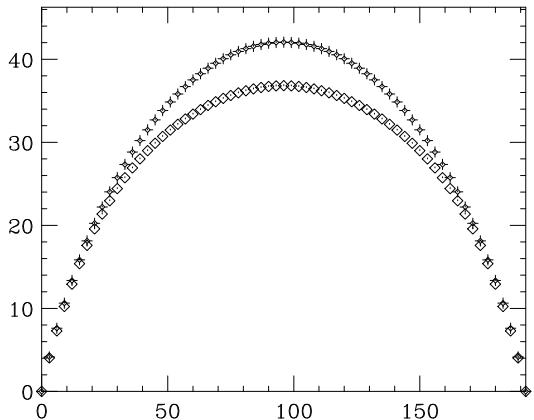,angle=90,width=200pt}
\caption{ As in fig. 1 for $V=4^3\times 20$.}
\label{fig:2}
\end{figure}
\section*{Finite coupling}
At non zero $\beta$ the pure gauge part of the action couples 
the Polyakov lines among themselves and any trivial
factorization in the partition function is not possible any more.
The partition function can be written as the product of the pure gauge 
partition function times the $\beta=0$ partition function times a
factor $R$:
\be\la{6}
R = \frac
{
\int [dU] e^{-\beta S_G(U)}\prod_{i=1}^{V_s}\left( 1 + 
\frac{c^2 Tr L_i + c Tr L_i^*}{c^3 + 1} \right)
}
{
\int [dU] e^{-\beta S_G(U)}
}
\ee
From the previous section we conclude that, in the zero temperature case,
the only possible contribution of the phase comes from (\ref{6}).
For large temporal extent $c$ approaches 0 or $\infty$ and in both cases 
the absolute value of all the factors in the numerator of (\ref{6})
is upper and lower bounded by a finite number.
This implies that $R$ behaves at most exponentially with the spatial
lattice volume $V_s$ except for $\mu=\log 2m$.
When this happens the phase is clearly not relevant and the free energy 
density reduces to the pure gauge contribution plus the infinite coupling 
term: we get again a first order saturation transition at $\mu_c=\log 2m$
for all the values of $\beta$.

At finite temperature the scenario is completely different.
The non analyticity in the $\Z(\beta=0)$ disappears and expression 
(\ref{6}) can also give a finite contribution in the free energy density.
Even if an exact evaluation of $R$ is not possible this points
toward a smooth crossover from the zero density to the saturation
regime as soon as $L_t$ is considered finite in the thermodynamic limit.
\begin{figure}[!t]
\psrotatefirst
\psfig{figure=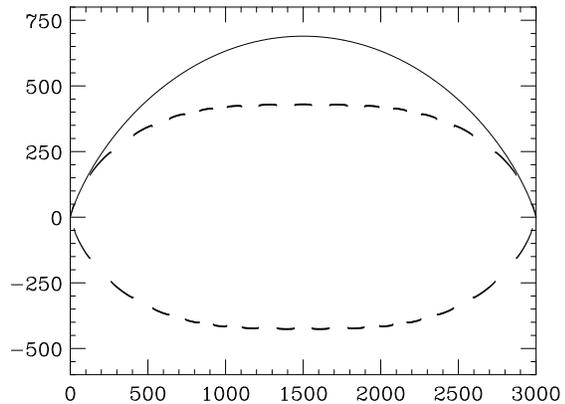,angle=90,width=210pt}
\caption{As in fig. 1: analytic (uppermost continuous line) and numerical  
results for $V=10^3\times 4$.}
\label{fig:3}
\end{figure}


\begin{thebibliography}{9}
\bibitem{bar1} I.M. Barbour, S.E. Morrison, E.G. Klepfish, J.B. Kogut, 
M.P. Lombardo, Phys. Rev. D56, 7063 (1997); Nucl. Phys. Proc. Suppl. 
60A (1998) 220.

\bibitem{noi1} R. Aloisio, V. Azcoiti, G. Di Carlo, A. Galante, A.F. Grillo, 
Phys. Lett. B428 (1998) 166; hep-lat/9804020 to appear in Phys. Lett. B.

\bibitem{step} M.A. Stephanov, Phys. Rev. Lett. 76 (1996) 4472.

\bibitem{noi2} R. Aloisio, V. Azcoiti, G. Di Carlo, A. Galante, A.F. Grillo, 
Nucl. Phys. Proc. Suppl. 63 (1998) 442.

\bibitem{tous} T.C. Blum, J.E. Hetrick, D. Toussaint,
Phys. Rev. Lett. 76 (1996) 1019.

\end{thebibliography}
\end{document}